\documentclass[aps,prb,preprint,groupedaddress,showpacs]{revtex4-1}
\usepackage{graphicx}
\usepackage{amsmath}

\begin{document}

\title{Landau levels, edge states and magneto-conductance in GaAs/AlGaAs core-shell nanowires}

\author{Miquel Royo}
\email{miguel.royovalls@unimore.it}
\author{Andrea Bertoni}
\email{andrea.bertoni@nano.cnr.it}
\affiliation{CNR-NANO S3, Institute for Nanoscience, Via Campi 213/a, 41125 Modena, Italy}
\author{Guido Goldoni}
\email{guido.goldoni@unimore.it}
\affiliation{Department of Physics, Informatics and Mathematics, University of Modena and Reggio Emilia, Italy}
\affiliation{CNR-NANO S3, Institute for Nanoscience, Via Campi 213/a, 41125 Modena, Italy}

\date{\today}

\begin{abstract}
Magnetic states of the electron gas confined in modulation-doped core-shell nanowires are calculated for a transverse field of arbitrary strength and orientation. Magneto-conductance is predicted within the Landauer approach. The modeling takes fully into account the radial material modulation, the prismatic symmetry and the doping profile of realistic GaAs/AlGaAs devices within an envelope-function approach, and electron-electron interaction is included in a mean-field self-consistent approach. Calculations show that in the low free-carrier density regime, magnetic states can be described in terms of Landau levels and edge states, similar to planar two-dimensional electron gases in a Hall bar.
However, at higher carrier density the dominating electron-electron interaction leads to a strongly inhomogeneous localization at the prismatic heterointerface. This gives rise to a complex band dispersion, with local minima at finite values of the longitudinal wave vector, and a region of negative magneto-resistance.
The predicted marked anisotropy of the magneto-conductance with field direction is a direct probe of the inhomogeneous electron gas localization of the conductive channel induced by the prismatic geometry.

\end{abstract}


\maketitle

\section{Introduction\label{introduction}}

Radially modulated semiconductor heterostructures, realized from core-(multi)shell nanowires (CSNWs), \cite{Morral08,keplingerNL09,heigoldtJMC09,Sladek10,stormNNANO2012} offer new perspectives in quantum electronics.\cite{Lieber07} Several crucial steps have been taken toward the realization of high-mobility devices based on this new class of nano-materials and their integration.\cite{IEEE2011} Single-crystal, defect-free cores using several III-V's,\cite{Shtrikman09a,Shtrikman09b} selective \emph{radial} doping,\cite{Tomioka10} high quality interfaces,\cite{Jiang2012} and integration with Si substrates\cite{Tomioka2012} have been realized.

\begin{figure}[ht]
\includegraphics{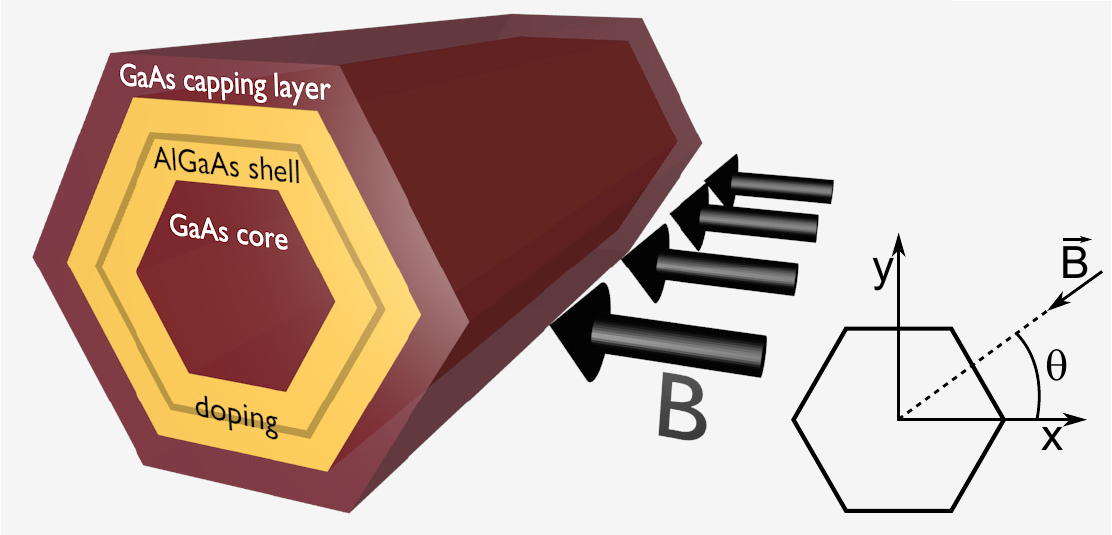}
\caption{Schematics of the prototypical radial heterojunction studied in this work. Remote doping in the middle of the AlGaAs layer results in an electron gas at the inner GaAs/AlGaAs interface. A transverse uniform magnetic field is applied. Inset: axis definitions and angle $\theta$ formed by the magnetic field.}
\label{fig1}
\end{figure}

Figure \ref{fig1} show the schematics of a prototypical GaAs/AlGaAs radial heterojunction. A GaAs nanowire, which typically grows along the [111] direction radially exposing the six \{110\} facets, is epitaxially overgrown by an AlGaAs shell,\cite{Spirkoska09} including a doping layer,\cite{Spirkoska11} and a GaAs capping layer, which protects the AlGaAs layer from oxidation.\cite{Li2011} Surface states of the outer GaAs layer, which lie about the midgap energy, easily deplete the outer layers of the structure, and an electron gas may form at the inner GaAs/AlGaAs hetero-interface.\cite{Sladek10}

Such radial modulation-doped heterojunctions can host a high-mobility electron gas, similarly to high-mobility two-dimensional electron gases (2DEGs), but wrapped around the core. However, due to the prismatic shape of the core and to electron-electron interaction, in a CSNW channel the electron gas will not be uniformly distributed at the heterointerfaces. Mean-field calculations \cite{Bertoni2011, wongNL11} \emph{at zero magnetic field} show that the electron gas distribution strongly depends on the free charge density.\cite{Tomioka2012} Three regimes can be identified\cite{Bertoni2011} for the conductive channel in GaAs CSNWs: i) a low density regime, with the electron gas cylindrically distributed in a quasi-1D channel into the GaAs core; ii) an intermediate density regime, with the central region of the core depleted and the electron gas mainly localized on a cylindrical surface next to the inner heterointerface; iii) a high density regime with the electron gas preferentially localized into a set of
quasi-1D channels, strongly tunnel coupled, and concentrated at the edges between different facets of the hexagonal core to maximize the inter-electron distance.
Interestingly, electronic states showing localization patterns in a similar fashion to this last regime have been recently demonstrated in two separate works on radially heterostructured hexagonal NWs,~\cite{MartinezNL12,FickenscheNL13} both combining optical measurements with theoretical simulations. In these studies, however, the strong localization effects arise from the spatial confinement occurring in narrow coaxial quantum wells.
Conductive channels in CSNWs are very sensitive to an external field and may be reshaped by an external gate, an important aspect, e.g., for gate-all-around (GAA) FET design.\cite{Tomioka2012, Kim2010}

Magnetic states in wrapped heterojunctions have not been investigated so far. Studies of the magnetic states on a cylindrical surface\cite{Ajiki93,Ferrari08B,Bellucci10} have been extended to prismatic surfaces.\cite{FerrariNL} These qualitative studies allow to highlight the influence of local topology on magnetic states in, e.g., a narrow wrapped around quantum well,~\cite{FickenscheNL13} but do not apply to doped heterojunctions, since electron-electron interaction is completely neglected. Unfortunately, no magneto-transport experiment in the Quantum Hall regime in CSNWs is available to date.

In this paper, we report self-consistent field calculations of realistic CSNWs subjected to a transverse magnetic field in the Quantum Hall regime. Our target systems are remotely doped GaAs/AlGaAs single radial heterojunctions, as the one shown in Fig.~\ref{fig1}. Magnetic states are calculated in different charge density regimes and different field orientations. Calculations show that magnetic states can be described in terms of Landau levels and edge states, similarly to 2DEGs, only in the low free-carrier density regime. At higher density the dominating electron-electron interaction leads to a strongly inhomogeneous localization at the prismatic heterointerface, giving rise to a complex band dispersion. The ensuing negative magneto-resistance and the marked anisotropy of the magneto-conductance with respect to field direction are a direct probe of the inhomogeneous electron gas localization of the conductive channel.

\section{Method\label{method}}

Within the effective mass, single parabolic band approximation, the  Hamiltonian in an external magnetic field reads
\begin{equation}
\hat{H}=\frac{1}{2}\left(\hat{\mathbf{P}} - e\, \mathbf{A}(\mathbf{R})\right) \frac{1}{m^*(\mathbf{R})}
\left(\hat{\mathbf{P}} - e \, \mathbf{A}(\mathbf{R})\right) + E_C(\mathbf{R}) - e\, V(\mathbf{R}).
\label{eq1}
\end{equation}
\noindent where $\mathbf{R}=(x,y,z)$, $\hat{\mathbf{P}}$ is the conjugate momentum operator, $\mathbf{A}(\mathbf{R})$ is the magnetic vector potential, and $m^*(\mathbf{R})$ is the position dependent, isotropic electron effective mass. $E_C(\mathbf{R})$ and $V(\mathbf{R})$ are, respectively, the local conduction band edge and the self-consistent potential generated by other electrons and the fully-ionized, static donors.


To describe a transverse magnetic field, namely, perpendicular to the nanowire axis, we employ the gauge
$\mathbf{A}(\mathbf{R}) =  B \cdot (0,0,\Theta(\mathbf{R}))$, with $\Theta(\mathbf{R})$ defined as
$\Theta(\mathbf{R})= y \cos\theta - x \sin\theta$.
Different orientations of the transverse field are obtained by the appropriate azimuthal angle $\theta$, which is measured with respect to the $x$ axis (see Fig.~\ref{fig1}). $\theta=0$ corresponds to a field oriented \emph{along a maximal diameter} of the hexagonal core, while $\theta=30^\circ$ corresponds to a field which is \emph{perpendicular to a facet} of the nanowire.

A uniform magnetic field in the direction transverse to the CSNW does not break the translational invariance along the wire axis, which is taken along the $z$ direction. 
Therefore, the 3D electron envelope function $\Psi_n(\mathbf{R})$ can be factorized in a 1D plane wave and a 2D envelope function, $\Psi_{n,k}(\mathbf{R})= e^{ikz} \phi_{n,k}(\mathbf{r})$, and labeled by the the principal quantum number, $n$, and the electron momentum along the $z$ direction, $k$, with $\mathbf{r}=(x,y)$.
By inserting $\Psi_{n,k}$ into the Hamiltonian, Eq.~(\ref{eq1}), the following equation for $\phi_{n,k}(\mathbf{r})$ is obtained:
\begin{equation}
\left\lbrace -\frac{\hbar^2}{2}\nabla_{\mathbf{r}}\left[\frac{1}{m^*(\mathbf{r})}\nabla_{\mathbf{r}}\right]
+ \frac{1}{2}m^*(\mathbf{r})\omega_c(\mathbf{r})^2\left(\Theta(\mathbf{r})-\Theta_0\right)^2 + E_C(\mathbf{r})
-e\,V(\mathbf{r})\right\rbrace \phi_{n,k}(\mathbf{r})= \epsilon_{n,k} \phi_{n,k}(\mathbf{r}).
\label{eq2}
\end{equation}
\noindent Here, $\omega_c$ is the cyclotron frequency, $\omega_c(\mathbf{r})=e\,B/m^*(\mathbf{r})$,
and $\Theta_0=k\,l_B^2$, with $l_B=\sqrt{\hbar/e\,B}$ being the magnetic length. Hence, in this gauge
the transverse magnetic field amounts to an effective parabolic potential lying in the plane defined by the field direction and the normal to wire axis. $\Theta_0$ is the vertex of the harmonic potential, which is displaced from the axis of the hexagonal section by $k\,l_B^2$ along the direction perpendicular to the field. This gauge is equivalent to the Landau gauge which in a planar Hall bar geometry gives rise to LLs in the bulk of the bar and ESs near to the boundaries.\cite{MacDonald84} Here, however, there are no electrostatically defined edges. In fact, lateral confinement in the direction perpendicular to the field results from charge redistribution in the complex three-dimensional structure, which has to be calculated self-consistently.

In zero magnetic field, subbands are parabolic, $\epsilon_{n,k}= \epsilon^0_{n}+\hbar^2 k^2/2 m^*$, with $\epsilon^0_{n}$ the energy at $k=0$, and $\phi_{n,k}(\mathbf{r})$ is $k$-independent. However, in the presence of the magnetic term the subbands develop non trivial dispersions, strongly dependent on the self-consistent potential entering in Eq.~(\ref{eq2}), as we shall see in the next section.

We solve Eq.~(\ref{eq2}) for magnetic fields of arbitrary strength and orientation at several $k$-points on a uniform grid in $[-k_{max}, k_{max}]$, with $k_{max}$ fairly above the Fermi wave vector. Then, the electron density is obtained from
\begin{equation}
n(\mathbf{r})= 2\, \sum_n\int_{-k_{max}}^{k_{max}} \frac{dk}{2\,\pi}\, f(\epsilon_{n,k}-\mu,T)
\left|\phi_{n,k}(\mathbf{r})\right|^2,
\label{eq3}
\end{equation}

\noindent where the leading $2$ on the right-hand-side accounts for spin degeneracy, and $f(\epsilon_{n,k}-\mu,T)$ is the Fermi occupation for each ($k,n$) state given by

\begin{equation}
f(\epsilon_{n,k},\mu,T)=\frac{1}{1+e^{(\epsilon_{n,k}-\mu)/k_B\,T}},
\label{eq4}
\end{equation}

\noindent $\mu$, $T$ and $k_B$ being, respectively, the Fermi energy, temperature, and Boltzmann constant. The chemical potential $\mu$ is pinned by the surface states of the outer GaAs layer, and in our calculation is fixed exactly at midgap.

Once the electron density is determined, the Hartree electrostatic potential $V_H$ due to the free charge and the ionized impurities is obtained from Poisson equation
\begin{equation}
\nabla\varepsilon(\mathbf{r})\nabla V_H(\mathbf{r})=\frac{1}{\varepsilon_0}e(n(\mathbf{r})-\rho_D(\mathbf{r})).
\label{eq5}
\end{equation}
\noindent Here, $\rho_D(\mathbf{r})$ is the volume density of dopants, which are considered to be fully ionized, while $\varepsilon(\mathbf{r})$ and $\varepsilon_0$ are the position-dependent dielectric constant and the vacuum permittivity.
An exchange-correlation correction calculated is added to the electrostatic potential within the local density approximation .\cite{gunnarssonPRB76, andoJAP03} We checked that its contribution does not exceed few percent of the mean-field potential.\cite{nota-sdft}

Equations (\ref{eq2}) and (\ref{eq5}) are numerically integrated iteratively, with Dirichlet boundary conditions, through a box integration method on a triangular grid with hexagonal elements.
The latter discretization, having the same symmetry of the integration domain, does not introduce numerical artifacts at the boundaries. No spatial symmetry is imposed, however. The original $D_{6h}$ symmetry of the Hamiltonian is reduced to $C_2$ by the transverse field and no degeneracies are therefore to be obtained in general.

The whole procedure is iterated until self-consistency is reached, which we consider to occur at the particularly strict condition that the relative variation of the charge density between two consecutive iterations is lower than 0.001 at any point of the discretization domain.
The simulation procedure has been checked against available data for planar 2DEG devices, as MOS capacitors. \cite{DattaQT2005}
We refer the reader to Ref.~\onlinecite{Bertoni2011} for more technical details about the numerical approach.

Once the convergence is achieved, energies and subbands occupations allow to estimate the
ballistic conductance of the NW by means of the linear-response Landauer formula,
\begin{equation}
G= \frac{e^2}{h}\sum_n \int_{{\cal B}_n} -\frac{\partial f(E-\mu,T)}{\partial E} \, dE.
\label{eq6}
\end{equation}
\noindent where the integral is performed on each (nonparabolic) energy band ${{\cal B}_n}$, from $E(-k_{max})$ to $E(k_{max})$, and gives a significant contribution only around the crossings of the band with the Fermy energy $\mu$.

\section{\label{magstates}Magnetic states}

We next discuss magnetic states of the prototypical GaAs CSNW shown in Fig.~\ref{fig1} for different charge density regimes, obtained changing the doping concentration in the simulation.\cite{Bertoni2011} We shall find it important to distinguish between two directions of the magnetic field, either perpendicular to a facet or along a maximal diameter (joining two opposite vertices).
In the next section we will show that the anisotropic response to the field is a direct evidence of the inhomogeneous localization of the electron gas.

Our reference sample~\cite{Sladek10} is a CSNW with a GaAs core having a facet-to-facet distance of 80 nm. The core is surrounded by a 50 nm thick $\mathrm{Al_{0.3}Ga_{0.7}As}$ shell and a 10 nm thick GaAs capping layer. The system is $n$-doped with a constant density of donors $\rho_D$, uniformly distributed in a 10 nm thick layer, placed in the middle of the $\mathrm{Al_{0.3}Ga_{0.7}As}$ shell. The Fermi energy is pinned by GaAs surface states at the middle of the gap, and is taken as the reference level for energies, i.e. $\mu=0$.

All calculations have been performed assuming a temperature of 4K. The position dependent material parameters employed in the calculations are
$m^*(\text{GaAs})=0.067$, $\varepsilon(\text{GaAs})= 13.8$, $E_c(\text{GaAs})=0.715$ eV,
and,
$m^*(\text{Al}_{0.3}\text{Ga}_{0.7}\text{As})=0.092$, $\varepsilon(\text{Al}_{0.3}\text{Ga}_{0.7}\text{As})= 12.24$, $E_c(\text{Al}_{0.3}\text{Ga}_{0.7}\text{As})=0.999$ eV.\

\begin{figure}[ht]
\includegraphics{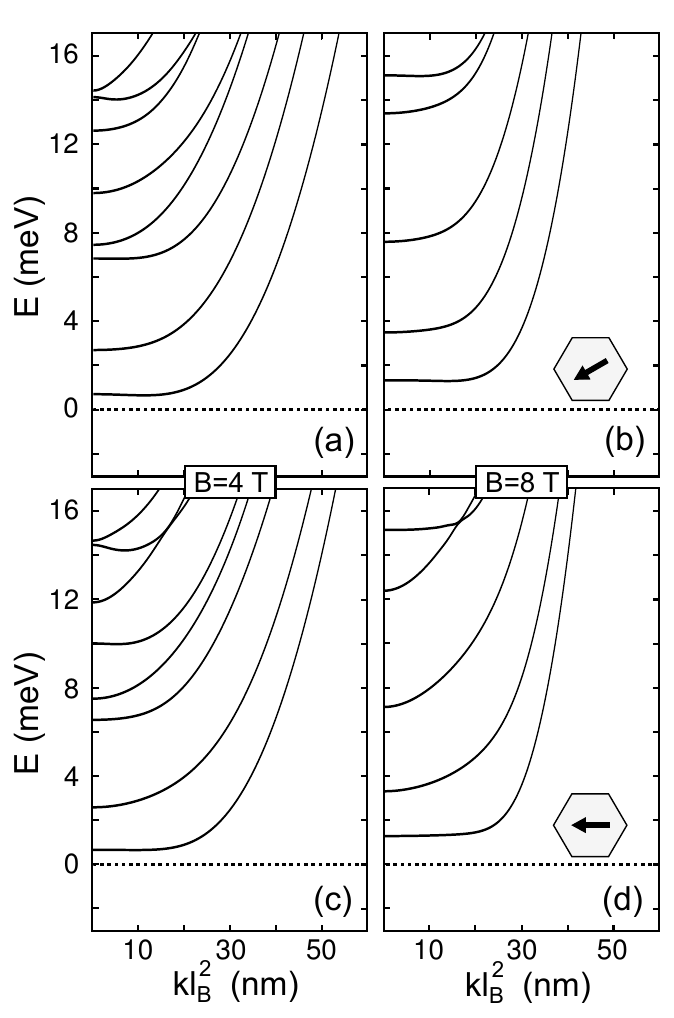}
\caption{Magnetic levels vs the in-wire momentum $k$ for a CSNW (see text) doped with $\rho_D=1.44\times 10^{18}\,\mathrm{cm^{-3}}$ at two representative fields, as indicated. The horizontal dashed
lines show the Fermi energy position. Top panels: field oriented perpendicular to the facets. Bottom panels: field oriented along a diameter.}
\label{fig2}
\end{figure}

\subsection{Low density regime\label{subsec:LDR}}

\begin{figure}[ht]
\includegraphics{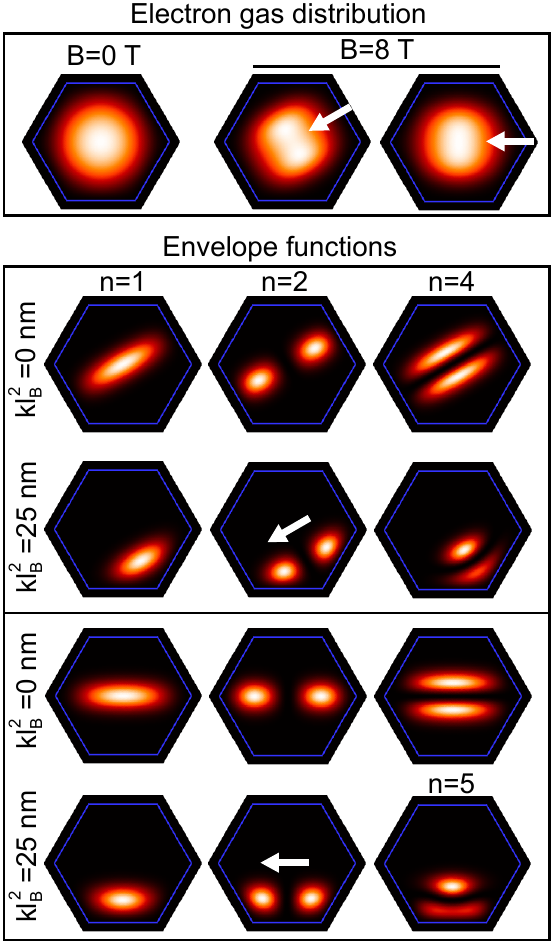}
\caption{Self-consistent charge density and envelope functions squared corresponding to magnetic states shown in Fig.~\ref{fig2}. Top panel: 2D self-consistent charge density at zero and finite magnetic field, as indicated, with the field oriented as shown by the arrows. Bottom panel: envelope functions squared at $B=8\,\mathrm{T}$ for selected
values of $(n,k) $, as indicated by labels. In each row, the field is oriented as shown by the white arrow in the center/bottom map.}
\label{fig3}
\end{figure}

We start the discussion from the low doping-density regime, with  $\rho_D=1.44\,\times
\, 10^{18}$ $\mathrm{cm^{-3}}$, which gives a linear free-electron density $0.014 \, \times \,10^7$ $\mathrm{cm^{-1}}$.  Figure~\ref{fig2} shows the subband dispersions in positive values of the in-wire momentum $k$ (the dispersion in negative $k$ is symmetric) at two representative values of the magnetic field, $B=4 \,\mathrm{T}$ (left) and $B= 8 \,\mathrm{T}$ (right), and for the two orientations of the field shown in the insets. Note that the lowest subband edge lies slightly above the Fermi energy (dashed line), so that the self-consistent charge density (Fig.~\ref{fig3}, top panel) is determined by a fractional thermal occupation of the lowest subband only. The charge density, which at zero field consists of a cylindrically symmetric quasi-1D channel extended over the GaAs core, is only slightly distorted by the field. It elongates in the direction normal to the field itself and develops two lobes.
This is a stronger effect with the field perpendicular  to the facets, as in this configuration the lateral electrostatic confinement is weaker.

The overall dispersion of the lowest subbands in Fig.~\ref{fig2} strongly resembles the LLs/ESs picture, familiar from planar 2DEGs in a Hall bar geometry.\cite{MacDonald84}
As the field is increased, the zero-field, parabolic dispersions at low $k$ gradually flatten, and form highly degenerate bands, similarly to LLs.
At the two representative fields shown in Fig.~\ref{fig2}, the flat region extends for a wide $k$ range. In fact, the magnetic length $l_B$, which at $4\,\mbox{T}$ and $8\,\mbox{T}$ is $\sim 13\, \mbox{nm}$ and $\sim 9\, \mbox{nm}$, respectively, is much smaller than the width of the GaAs core. At larger $k$, where $kl_B^2$ approaches the radius of the GaAs core, or, equivalently, the bottom of the magnetic harmonic potential approaches the lateral AlGaAs barriers by $\sim l_B$, the subbands bend up monotonically, similarly to ESs in a Hall bar.

The corresponding envelope functions are shown in the bottom panel of Fig.~\ref{fig3} for selected $(n,k)$ values. At $k=0$ electrons form LLs localized in the center of the structure, extending laterally, i.e., in the direction \emph{orthogonal} to the field, by about $l_B$. States with zero ($n=1,2$) or one ($n=4$) node in the \emph{vertical direction} (i.e. \emph{along} the direction of the field) correspond to the first and second LLs, respectively. At larger $k$ (see $kl_B^2=25~\mbox{nm}$ in Fig.~\ref{fig3}) states are displaced laterally, and correspond to ESs in a standard Hall bar (states with negative $k$ would be displaced in the opposite direction). Note that, depending on the field orientation, ESs localize either along a facet or an edge between two facets of the GaAs/AlGaAs interface.

The large thickness of the CSNW in the \emph{vertical direction} allows for several 1D subbands between successive LLs, similarly to an electron gas in a wide quantum well.\cite{Hayes91}
For instance, in Fig.~\ref{fig2}(b) the first three subbands correspond to the first LL.
They barely shift with the field [compare Figs.~\ref{fig2}(a,b)], since their energy is mainly determined by the vertical confinement.
On the other hand, the fourth subband corresponds to the second LL and its energy does shift linearly with the field.

Despite the anisotropic, hexagonal spatial confinement, the subband dispersions for the two field orientations are similar, which is consistent with the almost isotropic self-consistent electron density at $B=0$ shown in Fig.~\ref{fig3}. The subbands in Figs.~\ref{fig2}(c),(d) tend to be more dispersive than in Fig.~\ref{fig2}(a),(b), since for the former field orientation, as $k$ moves to large values, the states are pushed against a facet of the hexagonal core, and are continuously squeezed (along the field direction, see Fig.~\ref{fig3}) by the other two facets.

\subsection{Intermediate density regime \label{subsec:IDR}}

\begin{figure}[ht]
\includegraphics{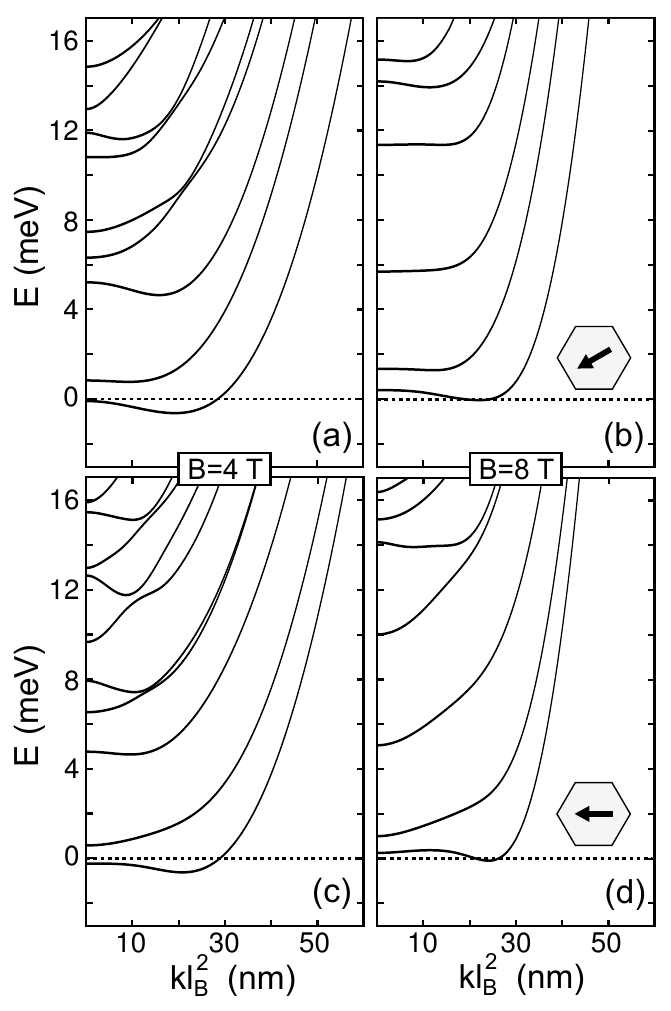}
\caption{Same as Fig.~\ref{fig2}, but with $\rho_D=1.5 \times 10^{18}\, \mathrm{cm^{-3}}$.}
\label{fig4}
\end{figure}

\begin{figure}[ht]
\includegraphics{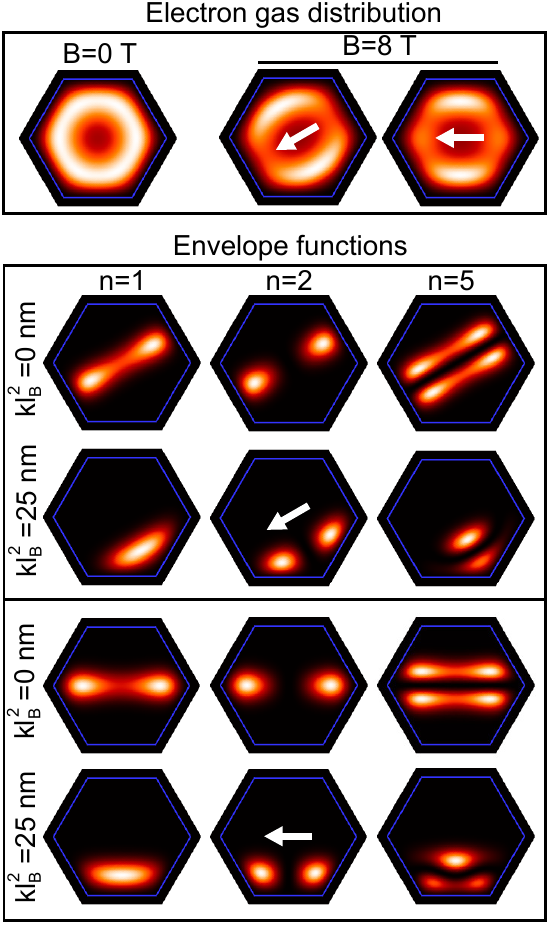}
\caption{Same as Fig.~\ref{fig3}, but corresponding to magnetic states in Fig.~\ref{fig4}.}
\label{fig5}
\end{figure}

We next consider a CSNW  with a slightly larger density of dopants, $\rho_D=1.5 \times 10^{18}\, \mathrm{cm^{-3}}$, corresponding to a linear electron density $\approx 0.096 \times 10^{7}\, \mathrm{cm^{-1}}$, about seven times larger than in the previous case.
The subband dispersions, and the self-consistent charge densities and envelope functions, are shown in Figs.~\ref{fig4} and~\ref{fig5}, respectively, at the same two representative fields of the previous case. The zero field self-consistent charge density, shown in Fig.~\ref{fig5}(top), consists of a hollow isotropic shell with finite thickness.\cite{Bertoni2011}
Therefore, it is expected that magnetic states resemble those of a cylindrical electron gas, which has been previously studied in the context of Carbon nanotubes\cite{Ajiki93} or cylindrical semiconductor systems.\cite{Ferrari08B} Indeed, although the overall subband dispersion still resembles the LLs/ESs structure, the lowest subband is not monotonic.  This type of dispersion is in agreement with single-particle calculations of the electron gas in the 2D surface of a cylinder.\cite{Ferrari08B}
The energy minimum corresponds to wavevectors such that electronic states may localize on the flanks of the structure, with respect to the magnetic field, i.e., in the regions where the normal component of the field, and therefore the field contribution to the energy, vanishes locally. There, the residual vertical component of the field changes sign and, therefore, acts as a restoring force, keeping electrons bound to quasi-1D channels. Semiclassically, these correspond to twisting or oscillating orbits,\cite{Ferrari08B,Bellucci10} and add to the usual ESs or skipping orbits typical of planar 2DEGs.\cite{Beenakker88}

Having two density of states singularities in the lowest subband, one at $k=0$ and one at a finite $k$, the Fermi energy is pinned to one of the two, depending on the field intensity. Note that the $k=0$ states are localized on top or bottom of the structure with respect to the field direction, while states near to the minimum at $k\neq 0$ are localized on one side (which one depending on the sign of $k$).
Accordingly, when the Fermi energy is pinned at $k=0$ the charge density (not shown) is nearly isotropic and similar to the zero-field case, while when it is pinned at $k\neq 0$, the electron density is concentrated in two lobes parallel to the field  direction (see Fig.~\ref{fig5}, top panel), each being composed of states which are propagating in the direction of $k$ or in the opposite direction.

The two lowest subbands are now nearly degenerate, since for this larger density the charge is pushed on opposite sides by the electron-electron interaction, similarly to a large quantum well,\cite{Hayes91} forming a symmetric/anti-symmetric (SAS) pair (see Fig.~\ref{fig5}, states with $n=1,2$). The pair is split in energy as $k$ moves from $k=0$ since, for these ESs, the vertical confinement is stronger and the tunneling energy is enhanced. Note also that, despite the apparent cylindrical symmetry at zero field, the response of the system to the magnetic field is more anisotropic than in the previous, low density case, particularly for large fields and higher subbands, as can be seen comparing the top and bottom panels in Fig.~\ref{fig4}.

\subsection{High density regime \label{subsec:HDR}}

\begin{figure}[ht]
\includegraphics{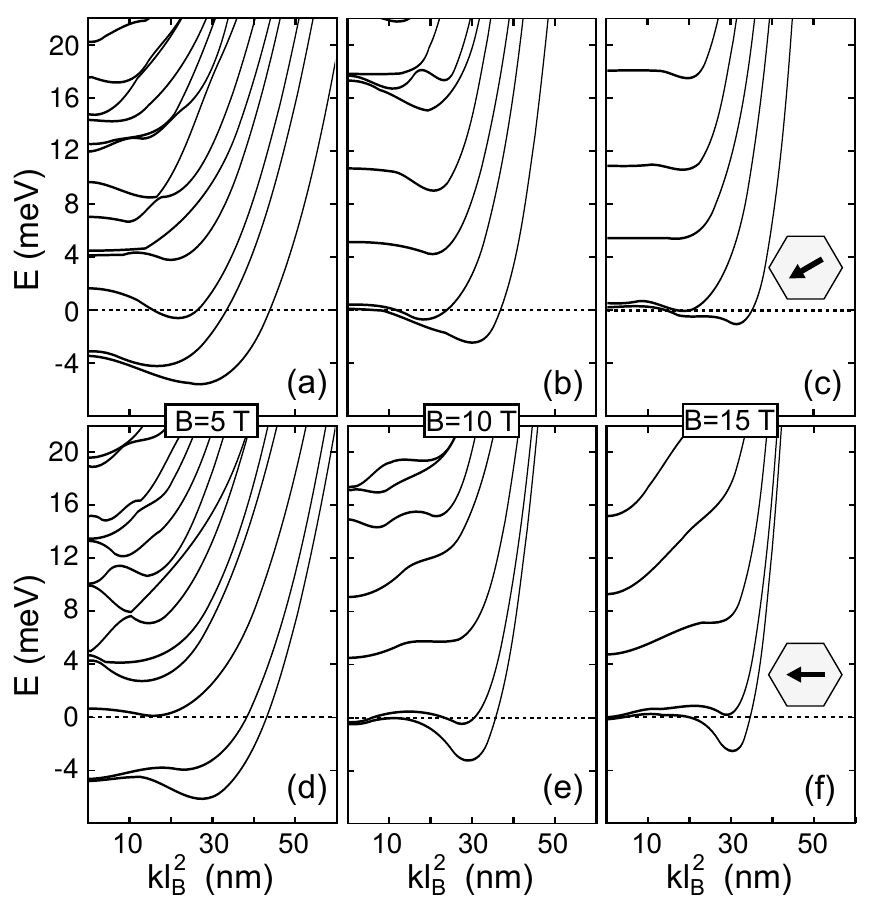}
\caption{Same as Fig.~\ref{fig2}, but with $\rho_D=1.7 \times 10^{18}\, \mathrm{cm^{-3}}$.}
\label{fig6}
\end{figure}

\begin{figure}[ht]
\includegraphics{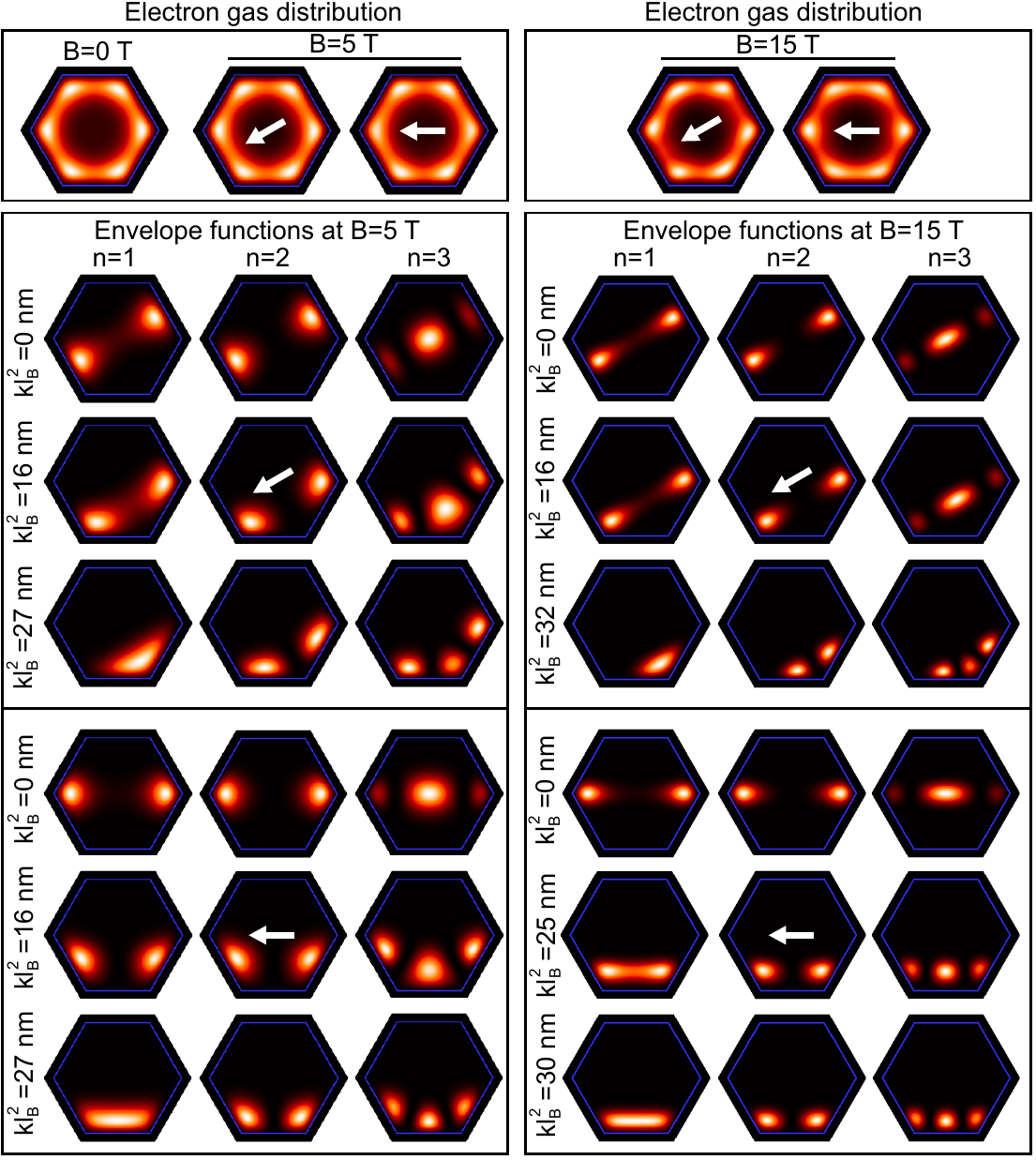}
\caption{Same as Fig.~\ref{fig3},  but corresponding to electron states in Fig.~\ref{fig6}. Charge densities and squared envelope functions are shown for both $B = 5 \mathrm{T}$ and $B = 15 \mathrm{T}$. \mathfrak{}}
\label{fig7}
\end{figure}

Finally, we consider the case of a NW with a density of dopants $\rho_D=1.7 \times 10^{18}\, \mathrm{cm^{-3}}$, corresponding to a linear electron density $0.425 \times 10^{7}\, \mathrm{cm^{-1}}$, about four times larger than the previous case.
This is the more complex situation, but possibly the most relevant one from the point of view of high-mobility transport experiments.
We show in Figs.~\ref{fig6} and~\ref{fig7} the subband dispersions and the self-consistent charge distributions and envelope functions, respectively.
Coulomb contribution is now dominating the Hamiltonian, and in this regime the zero field electron gas concentrates at the edges between two facets of the GaAs hexagonal core (Fig.~\ref{fig7}, top-left panel) clearly exposing the $\mathrm{D_{6h}}$ symmetry of the CSNW. A transverse field further reduces the symmetry, although this is mainly observed in the envelope functions, while the self-consistent electron density is barely affected by the field.

Let us now discuss separately the two field configurations, and first discuss the situation with the magnetic field applied perpendicularly to the facets.
Because of the strong localization at the GaAs/AlGaAs interfaces, due to electron-electron interaction, the lowest subbands form a SAS pair which is nearly degenerate at small $k$ (see Fig.~\ref{fig6}).
As the wavevector increases, both subbands bend down, corresponding to the formation of laterally confined states described also in the previous case, while SAS states with larger $k$ split, since the magnetic fields confines the wave functions laterally.
We show how this happens at the two representative fields $5\,\mbox{T}$ and $15\,\mbox{T}$ in Fig.~\ref{fig7}, corresponding to Fermi energy pinning at the third subband and at the lowest SAS doublet, respectively. At the lower field, the situation is similar to lower densities, with LLs smoothly evolving in ESs through one minimum, because the magnetic confinement has a larger lengthscale than the electrostatic confinement at the edges. In contrast, at 15 T, although the states have a more flat dispersion, which is expected because of the larger field, they go through a double minimum. This is because at a specific $k$ value the magnetic potential is centered exactly at the position of the self-consistent minima at the edges between the facets and with a similar confinement length (see $k\,l_B^2=16$ nm in Fig.~\ref{fig6}). At larger $k$, ESs develop, as in the previous cases. The excited subbands still develop LL-like flat band dispersion because they are spread over the center of the GaAs core, in a
situation similar to the subbands illustrated in Fig.~\ref{fig2}. Correspondingly, the magnetic field applied perpendicular to the facets slightly favors localization in the two edges along the orthogonal direction, but since the field adds to the strong self-consistent field here, this is only a minor effect (Fig.~\ref{fig7}, top right).

We next consider a magnetic field applied in the direction of a maximal diameter of the hexagonal section. As shown in Fig.~\ref{fig6} (bottom panels), the flat band dispersion of a LL is now hardly observed even in higher bands, since for this orientation, as $k$ increases the vertical (parallel to the field) spatial confinement decreases continuously (see Fig.~\ref{fig7}, bottom panels).
The two lowest subbands, which are nearly degenerate at $k=0$, bend up with $k$ and relax the degeneracy for the same reason, finally developing a minimum at finite $k$. The corresponding evolution of the wave functions with $k$ is illustrated in the bottom panels of Fig.~\ref{fig7}.
At large field the dispersion is almost flat, until the states develop a minimum, which is much deeper than in the previous field orientation. Indeed, this corresponds to states which are delocalized along a facet.

\begin{figure}[ht]
\includegraphics{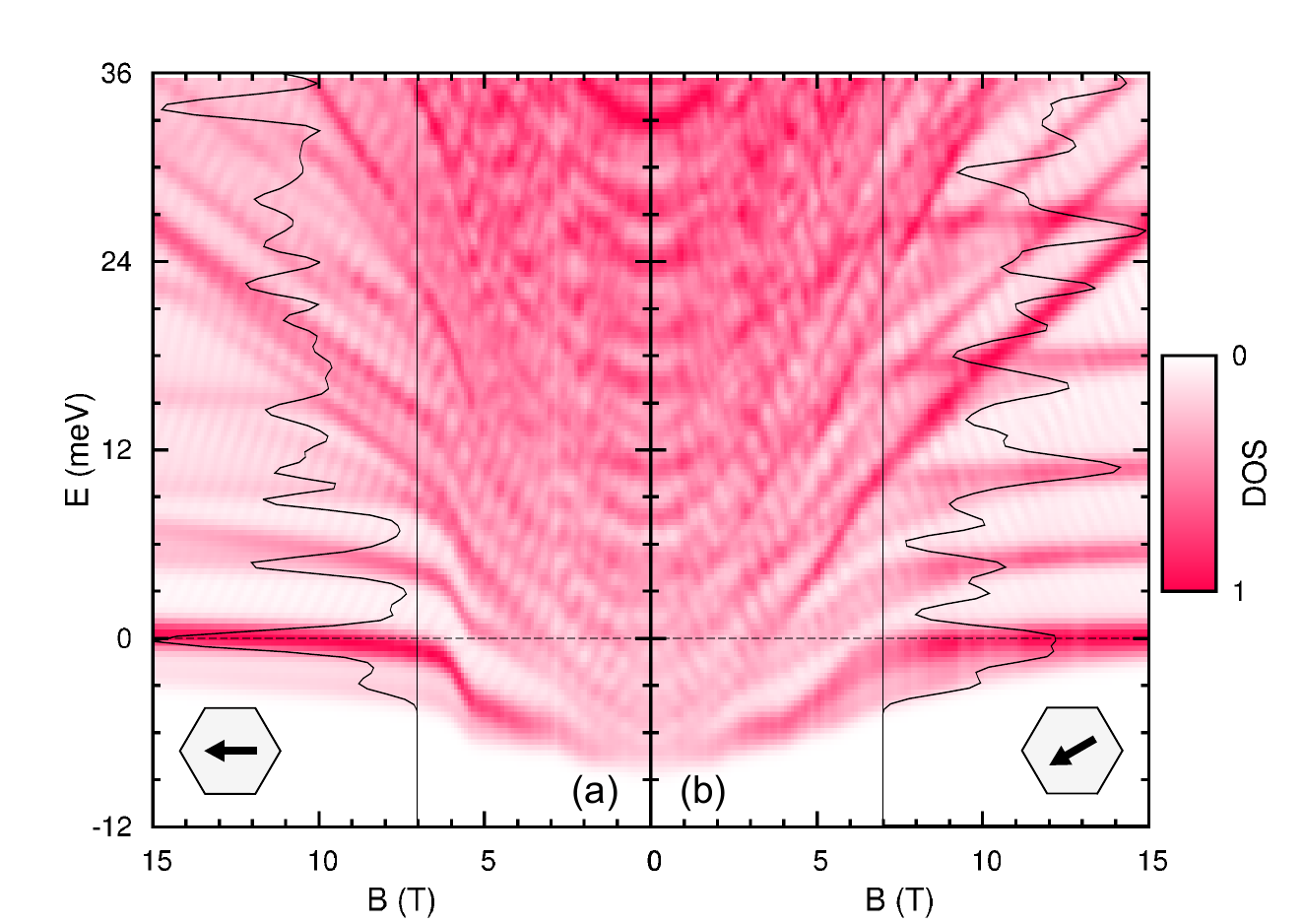}
\caption{Color intensity: normalized DOS vs transverse magnetic field intensity, with the field oriented as indicated in the insets. The Fermi energy is shown by a horizontal dashed line. In the left and right panels, a profile of the DOS at a specific field, indicated by a vertical line, is shown.}
\label{fig8}
\end{figure}

In Fig.~\ref{fig8} we show the density of states of the NW with a density of dopants $\rho_D=1.7 \times 10^{18}\, \mathrm{cm^{-3}}$ for both field orientations. LLs are easily identified at high fields as narrow peaks shifting linearly with the field. Since the Fermi energy is fixed by surface states at midgap, the lowest level is actually flat and pinned at the Fermi energy at sufficiently large field (correspondingly, as the field increases, the bottom of the self-consistent potential moves linearly at lower energies).
Each LL comes as a set of parallel bands arising from the vertical confinement. However, the DOS cannot be described only in terms of highly degenerate LLs. This is exemplified by the two DOS profiles shown in Fig.~\ref{fig8} at selected magnetic fields, which show a rich structure on top of the LL peaks, which is due to the dispersive states localized on the flakes of the structure.\cite{FerrariNL} In particular, one can recognize high DOS below the lowest LL at intermediate fields.

\section{Magneto-conductance}

We next focus on the magneto-conductance of a radial heterojunction. It is useful to recall that in a planar Hall bar or in high-mobility quasi-1D channels, as quantum wires~\cite{HewPRL09} or quantum point contacts,~\cite{MillerNAT07} the high-field longitudinal magneto-conductance typically shows a monotonic decrease with increasing field with quantized plateaus in units of $2e^2/h$ (for spinless electrons) at sufficiently low temperature, corresponding to field induced depopulation of the current-carrying ESs emerging from the highly degenerate LLs with monotonic dispersion.

\begin{figure}[ht]
\includegraphics[width=0.75\linewidth]{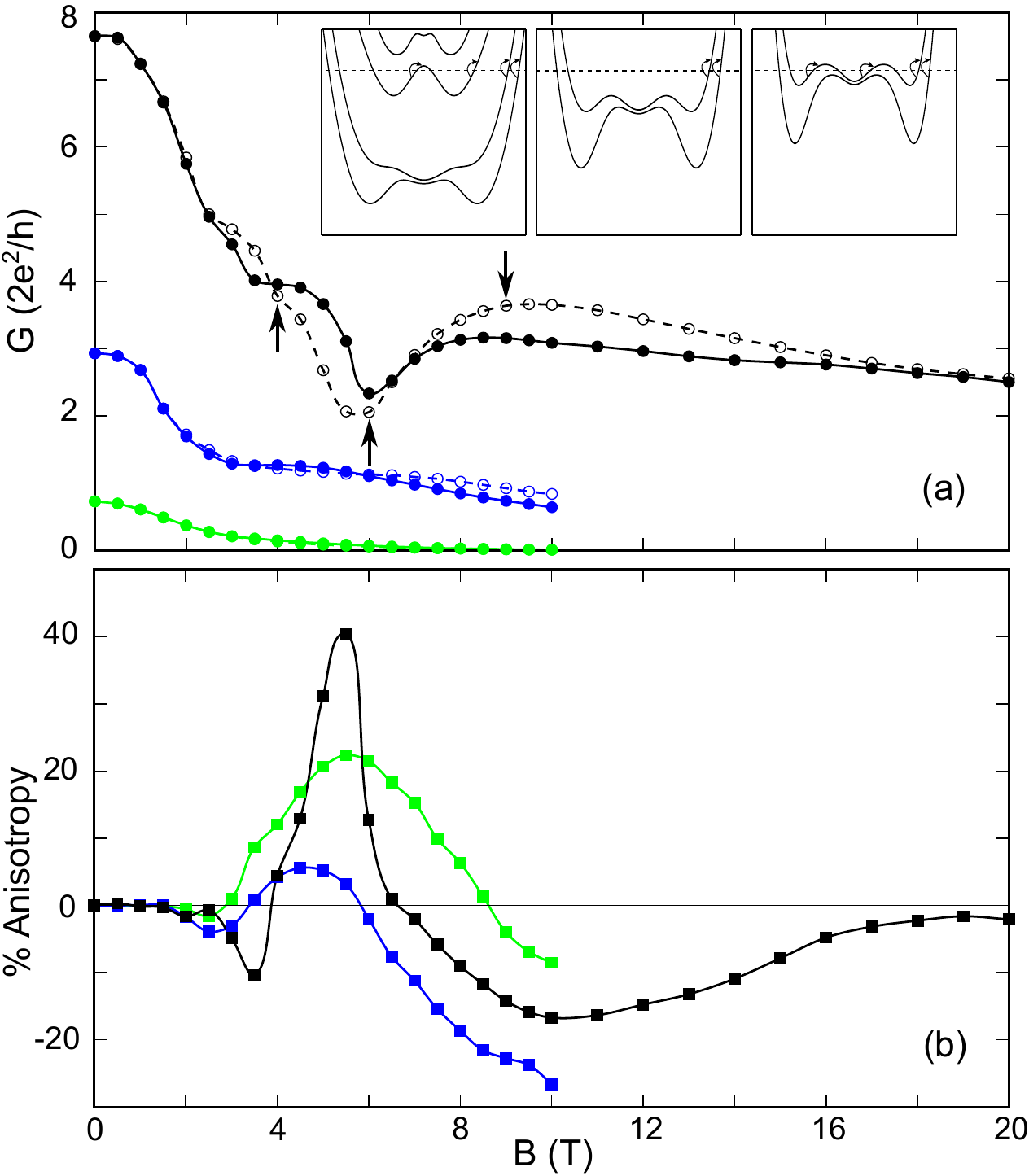}
\caption{(a) Magneto-conductance at three different doping densities: $\rho_D=1.7 \times 10^{18}\, \mathrm{cm^{-3}}$ (upper black curves), $\rho_D=1.5 \times 10^{18}\, \mathrm{cm^{-3}}$ (middle blue curves), and $\rho_D=1.44 \times 10^{18}\, \mathrm{cm^{-3}}$ (lower green curves). Solid lines: field oriented along a maximal diameter. Dashed lines: field oriented normal to a facet. Insets: magnetic states and intra-subband excitations contributing to the conductance at the three values of $B$ indicated by the arrows of the main plot, in the same order.
(b) Magneto-conductance anisotropy from Eq.~(\ref{eq6}).}
\label{fig9}
\end{figure}

Although CSNWs are quasi-1D systems, magnetic states have a complex subband structure and DOS which can be exposed in the peculiar magneto-conductance behavior. In Fig.~\ref{fig9}(a) we show the calculated magneto-conductance for the three doping densities considered before at $T=\,4\mathrm{K}$. For the lowest doping density the conductance is $G<1$ even at zero field due to the partial occupation of the first subband. As the magnetic field increases, the NW depletes completely and the conductance vanishes. At the intermediate doping density ($\rho_D=1.5 \times 10^{18}\, \mathrm{cm^{-3}}$), there are three occupied subbands at zero field. The second and third subbands are degenerate by symmetry at zero field, but when a finite magnetic field is applied, the degeneracy is removed and the two subbands subbands become consecutively depleted. As a result, the magneto-conductance rapidly reduces from $G\simeq3$ to $G\simeq1$. The large
plateau at $G\simeq 1$ correspond to the pinning of the Fermi level to the lowest subband (see, e.g., Fig.~\ref{fig4}). Finally, for the highest density of dopants there are eight subbands crossing the Fermi level at low fields. The magnetic field depletes the subbands up to $B\sim6$ T, where the magneto-conductance attains a minimum with $G\simeq2$.
Larger fields result in increased conductance, which finally falls to $G\simeq 2$ as the lowest subband is depleted.

The origin of the negative magneto-resistance is illustrated in the insets of Fig.~\ref{fig9}(a) (subbands are shown for field oriented along the diameter). At $B=4\, \mbox{T}$ (left inset) there are four states at the Fermi energy with a positive slope.
At $B=6\,\mbox{T}$ (center inset) the Fermi energy lies between the second and third subbands and only two channels are accessible, in a region with a low DOS (see Fig.~\ref{fig8}).
For large fields the conductance increases due to the non monotonic bending of the two lowest subbands, which results in additional channels at higher fields (as for $B\sim9\, \mbox{T}$, right inset).

The magneto-conductance also show substantial anisotropy. Note that the negative magneto-resistance region, shown in Fig.~\ref{fig9}(a) by the sample with larger doping, exists for both orientations of the field, and it is more pronounced when the field is oriented along a maximal diameter. The anisotropy
\begin{equation}
\frac{G_{1}-G_{2}}{(G_{1}+G_{2})/2}\times100\, ,
\end{equation}
where $G_1, G_2$ are the magneto-conductances for the field applied perpendicular to a facet and along a diameter, respectively, is shown in Fig.~\ref{fig9}(b). Note, in particular, that the magneto-conductance minimum is attained at a slightly different field for the two directions of the field, due to the different localization of the electron gas in different directions. Accordingly, the anisotropy changes sign in this region. Therefore, the change of sign of the anisotropy is a clear signature of the inhomogeneous localization of the electron gas in CSNWs.

\begin{figure}[ht]
\includegraphics[width=1.\linewidth]{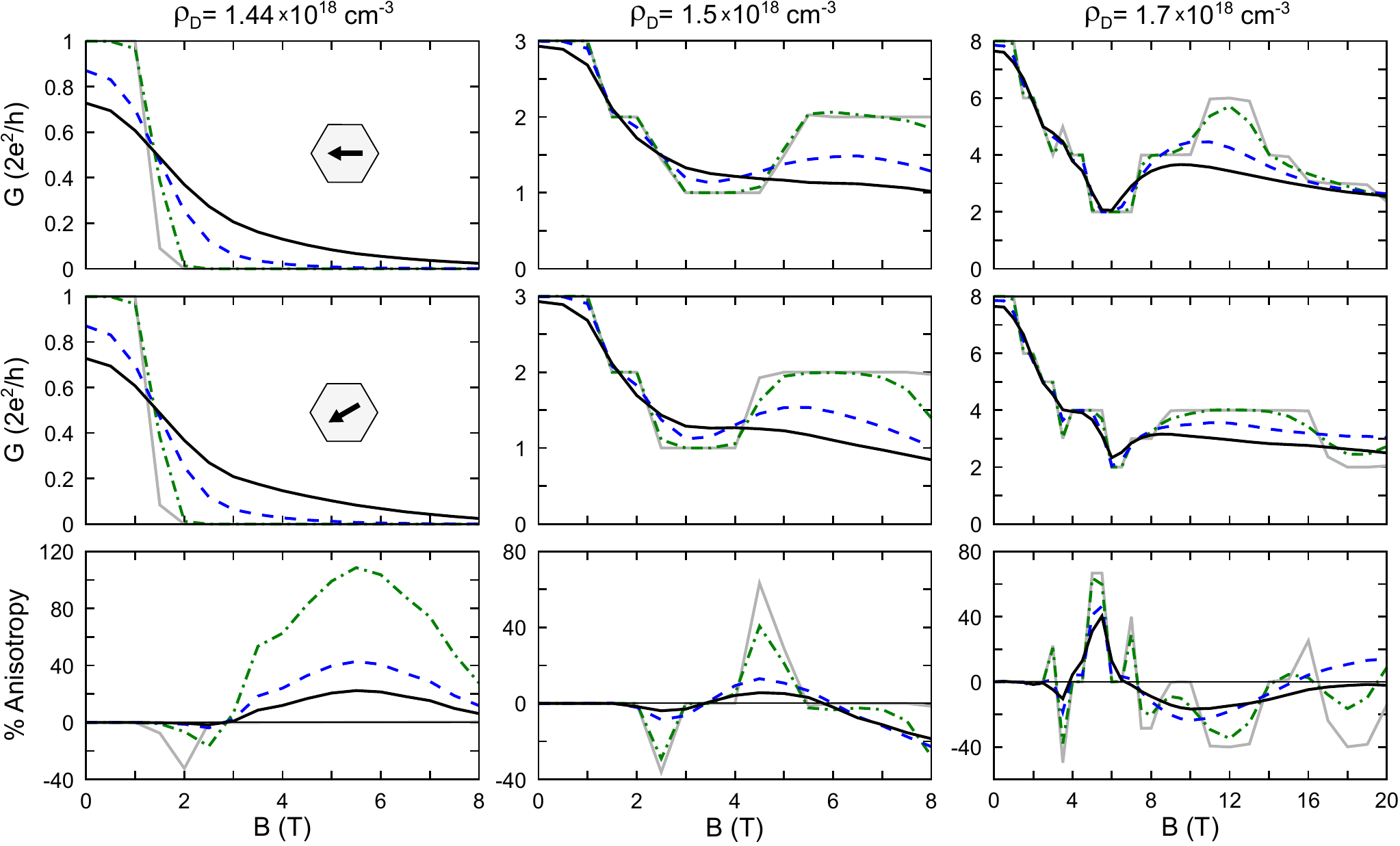}
\caption{Magneto-conductance at selected temperatures: $0.01 \,\mbox{K}$ (solid gray lines), $0.5\,\mbox{K}$ (dashed lines), $2\,\mbox{K}$ (dash-dotted lines), $4\,\mbox{K}$ (dotted lines). Top row: field oriented along a maximal diameter. Middle row: field oriented normal to a facet. Bottom row: magneto-conductance anisotropy. From left to right: the three columns display results for doping densities as indicated. }
\label{fig10}
\end{figure}

The dependence of the magneto-conductance on temperature is illustrated in Fig.~\ref{fig10}. The step-like profiles of the lowest,
$0.01$ K, temperature case (gray lines) ~\cite{HewPRL09,MillerNAT07,IhnatsenkaPRB08} are rapidly smoothed out by thermal broadening
(note that temperature enters here through the fractional Fermi occupations in Eq.~\ref{eq6}; we use the self-consistent subbands obtained
at $T=4\,\mbox{K}$ for all $T$ and check, in selected cases, that results are unaffected by this choice). However, anisotropy and negative magneto-resistance persists at larger temperatures.

We now note that the negative magneto-resistance regime is present (at sufficiently low temperature) both in the high and intermediate density regimes which, as discussed in the previous section, correspond to remarkably different electron gas localizations (compare Fig.~5 and 7, top panels). In both cases, the predicted increase of $G$ after a minimum originates in the bending of the lowest subbands. In the intermediate density regime two local minima emerge, one at positive and one at negative values of the in-wire momentum $k$ (see, e.g, section~\ref{subsec:IDR}), and arise from the fact that in a cylindrical geometry 
the vertical component of the field, and therefore the magnetic confinement energy, is minimum on the flanks of the cylinder. 
Indeed, similar predictions of negative magneto-resistance at low temperature have been reported for electron gases with tubular geometry.~\cite{magarill98JETP, TserkovnyakPRB06} 

However, we observe that the magnitude of 
the negative magneto-resistance in a NW in the intermediate density regime is of one unit of $G$ at the most, whereas in the high
density regime it can be of up to four units (see Fig.~\ref{fig10}, top-right panel). This is due to the wavier dispersion of the
subbands in the high density regime which allows more channels at the Fermi energy. As we have shown in section~\ref{subsec:HDR}, additional minima appear in the high density regime, and arise from the localization of magnetic states in the wells of the self-consistent potential, at the corners of the hexagonal core. Hence, the observation of a large negative magneto-resistance would be another signature of the electron gas localization in the edges of the NW.

\section{Conclusions}

In this work, we simulated a CSNW device through a self-consistent mean field procedure and determined the following.
First. Small variations of the doping density are able to modify the charge localization pattern from a 1D regime (electrons in the core) to a cylinder-like wrapped 2DEG (along the whole heterointerface), to a set of six coupled 1D channels (at the vertices of the hexagonal section of the heterointerface).
Second. The orthogonal magnetic field does not change much, at least qualitatively, the above localization patterns, but has a strong impact on the subband dispersions and on the density of states. Namely, novel minima at $k\neq 0$ appear, due to the competition between the magnetic length and the structure confinement. High DOS regions of the self-consistent subband structure pin to the Fermi level and a strong magnetic field enhances this effect due to the flattering of the bands at low $k$.
Third. Electron-electron interaction leads to strongly inhomogeneous localization and is responsible for the stability of the 2DEG also in presence of strong magnetic fields.
Fourth. The full inclusion of the prismatic cross-section in the numerical modeling of the CSNW is essential, since a cylindrical model, not taking into account the real sample symmetry, cannot reproduce the complex localization patterns and the different subband tailoring induced by the magnetic field.
Fifth. A regime of negative magneto-resistance is predicted in the case with high doping density, contrary to the case of planar 2DEGs or purely 1D systems. Furthermore, a substantial anisotropy is expected in the magneto-conductance, as the  magnetic field is applied along the direction joining two opposite vertices or perpendicular to two opposite facets. The last two effects are clear signatures of the inhomogeneous localization of the electron gas.

Since semiconductor nanowires are at the hearth of novel FET devices with a GAA patterning~\cite{Kim2010}, the analysis of carrier states and conductance is important also for applications.
For example, the effect of gate potential on the I-V characteristics of cylindrical GAA FET is strongly connected to the shape and position of the conduction channel. In CSNWs, the latter can be tailored with an external magnetic field and probed by magnetoconductance measurements, as we demonstrated.
We also note that current cylindrical models for the core quantum wire of GAA FETs may become soon inadequate, as its diameter moves from the micrometric to the nanometric scale.
Finally, a properly engineered CSNW can effectively move the free carriers of the FET channel closer to the gate electrode with respect to a bulk nanowire.  As a consequence, the effectiveness of the gate potential should be enhanced.

\bibliography{Bibliography}

\begin{thebibliography}{37}
\bibitem [{not()}]{nota-sdft}%
  \BibitemOpen
  \href@noop {} {}\bibinfo {note} {It has been shown that in the regime
  addressed in the present paper the spin-DFT and mean-field approaches give
  very similar results~\protect\cite{IhnatsenkaJPCM08} for the magnetic
  dependence of subbands.}\BibitemShut {Stop}%
\end{thebibliography}

\end{document}